\documentclass[aps,10pt,notitlepage,twocolumn,nofootinbib,superscriptaddress]{revtex4-2}

\usepackage{graphicx,color}
\usepackage{amsmath}
\usepackage{amssymb}
\usepackage{hyperref}
\usepackage[utf8]{inputenc}
\usepackage[english]{babel}
\usepackage{epsfig}
\usepackage{subfigure}
\usepackage{wasysym}
\usepackage{color,xcolor}
\usepackage{amsmath}
\usepackage{bm}
\usepackage{epsfig}
\usepackage{amsfonts}
\usepackage{dcolumn}
\usepackage{float}

\usepackage{comment}

\hypersetup{colorlinks=true, linkcolor=blue, citecolor=green}

\usepackage{dcolumn}
\usepackage{bm}
\usepackage{ifpdf}
\usepackage{hyperref}
\usepackage{float}
\usepackage{bm}
\usepackage{xcolor,color,graphicx,graphics}
\usepackage[OT1]{fontenc}
\usepackage{latexsym,amssymb,amsmath,amsfonts}
\usepackage{makeidx}
\usepackage{epsfig}
\usepackage{epstopdf}
\usepackage{mathrsfs}
\hypersetup{colorlinks=true, linkcolor=blue, citecolor=green}
\usepackage{enumerate}
\usepackage{xcolor}
 \usepackage{multirow}

\begin{document}

\title{Charged black hole solutions in $f(R,T)$ gravity coupled to nonlinear electrodynamics}

\author{Gabriel I. Róis } \email{fc54507@alunos.fc.ul.pt}
\affiliation{Instituto de Astrof\'{i}sica e Ci\^{e}ncias do Espa\c{c}o, Faculdade de Ci\^{e}ncias da Universidade de Lisboa, Edifício C8, Campo Grande, P-1749-016 Lisbon, Portugal}
\affiliation{Departamento de F\'{i}sica, Faculdade de Ci\^{e}ncias da Universidade de Lisboa, Edif\'{i}cio C8, Campo Grande, P-1749-016 Lisbon, Portugal}

    \author{José Tarciso S. S. Junior}
    \email{tarcisojunior17@gmail.com}
\affiliation{Faculdade de F\'{i}sica, Programa de P\'{o}s-Gradua\c{c}\~{a}o em F\'{i}sica, Universidade Federal do Par\'{a}, 66075-110, Bel\'{e}m, Par\'{a}, Brazill}

\author{Francisco S. N. Lobo} \email{fslobo@ciencias.ulisboa.pt}
\affiliation{Instituto de Astrof\'{i}sica e Ci\^{e}ncias do Espa\c{c}o, Faculdade de Ci\^{e}ncias da Universidade de Lisboa, Edifício C8, Campo Grande, P-1749-016 Lisbon, Portugal}
\affiliation{Departamento de F\'{i}sica, Faculdade de Ci\^{e}ncias da Universidade de Lisboa, Edif\'{i}cio C8, Campo Grande, P-1749-016 Lisbon, Portugal}

	\author{\\Manuel E. Rodrigues} \email{esialg@gmail.com}
\affiliation{Faculdade de F\'{i}sica, Programa de P\'{o}s-Gradua\c{c}\~{a}o em F\'{i}sica, Universidade Federal do Par\'{a}, 66075-110, Bel\'{e}m, Par\'{a}, Brazill}
\affiliation{Faculdade de Ci\^{e}ncias Exatas e Tecnologia, Universidade Federal do Par\'{a}, Campus Universit\'{a}rio de Abaetetuba, 68440-000, Abaetetuba, Par\'{a}, Brazil}

\author{Tiberiu Harko} \email{tiberiu.harko@aira.astro.ro}
\affiliation{Department of Physics, Babes-Bolyai University, 1 Kogalniceanu Street, 400084 Cluj-Napoca, Romania}
\affiliation{Astronomical Observatory, 19 Ciresilor Street, 400487, Cluj-Napoca, Romania}


\begin{abstract}
In this work, we investigate static and spherically symmetric black hole solutions in $f(R,T)$ gravity, where $R$ is the curvature scalar and $T$ is the trace of the energy-momentum tensor, coupled to nonlinear electrodynamics (NLED). 
To construct our solutions, we adopt a linear functional form, $f(R,T) = R + \beta T$. In the limit $\beta = 0$, the theory reduces to General Relativity (GR), recovering $f(R,T) \approx R$.
We propose a power-law Lagrangian of the form $\mathcal{L} = f_0 + F + \alpha F^p$, where $\alpha =f_0= 0$ corresponds to the linear electrodynamics case. Using this setup, we derive the metric functions and determine an effective cosmological constant. Our analysis focuses on specific cases with $p = 2$, $p = 4$, and $p = 6$, where we formulate analytic expressions for the matter fields supporting these solutions in terms of the Lagrangian as a function of $F$. Additionally, we verify the regularity of the solutions and study the structure of the event horizons.
Furthermore, we examine a more specific scenario by determining the free forms of the first and second derivatives $\mathcal{L}_F(r)$ and $\mathcal{L}_{FF}(r)$ of the Lagrangean of the nonlinear electromagnetic field. From these relations, we derive the general form of $\mathcal{L}_{\text{NLED}}(r)$ using consistency relations. This Lagrangian exhibits an intrinsic nonlinearity due to the influence of two constants, $\alpha$ and $\beta$. Specifically, $\alpha$ originates from the power-law term in the proposed Lagrangian, while $\beta$ arises from the assumed linear function $f(R,T)$.
The interplay of these constants ensures that the nonlinearity of the Lagrangian is governed by both $\alpha$ and $\beta$, rather than $\alpha$ alone. By imposing specific constraints, namely, setting $\alpha = 0$ and $\beta = 0$, the model reduces to the linear electrodynamics case, while still remaining consistent with the $f(R,T)$ gravity framework. This demonstrates the model's flexibility in describing both linear and nonlinear regimes.
\end{abstract}
\maketitle
\def\HMS{{\scriptscriptstyle{\rm HMS}}}

\tableofcontents

\section{Introduction}\label{sec1}

The theory of General Relativity (GR), introduced by Albert Einstein~\cite{Einstein:1916vd}, has captivated the scientific community since its inception. One of its most fascinating predictions emerged soon thereafter: the existence of black holes. In recent years, two key predictions of this groundbreaking theory have been spectacularly confirmed, reigniting interest in the study of some of the universe's most enigmatic phenomena.
One of the most recent of these confirmations occurred in 2015, when the LIGO and VIRGO collaborations made the first detection of gravitational waves~\cite{LIGOScientific:2016aoc, LIGOScientific:2017ync}. These ripples in spacetime, caused by violent astrophysical events, provided direct evidence of one of GR's most remarkable implications. Shortly thereafter, the existence of black holes was visually confirmed by the Event Horizon Telescope (EHT) collaboration. This team achieved a historic milestone by capturing the first image of the superheated plasma surrounding the black hole at the center of galaxy M87~\cite{EventHorizonTelescope:2019dse}. More recently, the EHT unveiled an image of Sagittarius A* (Sgr A*), the supermassive black hole at the heart of our own Milky Way galaxy~\cite{EventHorizonTelescope:2022wkp}.
These groundbreaking achievements have paved the way for a transformative era in gravitational physics, allowing us to explore the universe and its phenomena with unprecedented precision. By confirming GR’s predictions and revealing the extreme environments near black holes, these discoveries have opened new pathways to understanding the cosmos in ways that were once beyond imagination.

As GR encounters limitations in explaining certain phenomena, such as dark matter, cosmic inflation, and the late-time acceleration of the universe (attributed to dark energy), alternative approaches have emerged to deepen our understanding of these mysteries.
One prominent avenue involves generalizations of GR that replace the Einstein-Hilbert action with arbitrary functions of the Ricci scalar $R$, leading to the so-called $f(R)$ modified theories of gravity \cite{Nojiri:2006ri,Sotiriou:2008rp,DeFelice:2010aj,Clifton:2011jh}. A notable example is Starobinsky’s theory, which incorporates quantum corrections to inflationary models by introducing an $R^2$ term into the Einstein-Hilbert action~\cite{Starobinsky:1980te}.
Another significant extension is the inclusion of the trace of the energy-momentum tensor, $T$, in the gravitational action, giving rise to $f(R,T)$ gravity \cite{Harko:2011kv,Barrientos:2018cnx}. Here, $T$ represents the trace of the energy-momentum tensor $T_{\mu\nu}$. 
The presence of explicit nonminimal curvature-matter couplings result in a nonzero covariant derivative of the energy-momentum tensor \cite{Gonner:1984zx,Bertolami:2007gv,Bertolami:2008zh,Bertolami:2008ab,Harko:2010mv,Harko:2012ve,Harko:2012hm,Haghani:2013oma,Harko:2014gwa,Harko:2020ibn}. This leads to deviations from geodesic motion and the emergence of an extra force \cite{Gonner:1984zx,Bertolami:2007gv}.
Since its inception, $f(R,T)$ gravity has been applied across a wide range of contexts. These include studies in thermodynamics~\cite{Sharif:2012zzd}, analyses of energy conditions~\cite{Sharif:2012ce,Alvarenga:2012bt}, investigations of compact stars \cite{Moraes:2015uxq,Das:2016mxq,Maurya:2019hds,Maurya:2019sfm}, and gravastar solutions~\cite{Das:2017rhi,Yousaf:2019zcb}. Moreover, it has been extensively explored in cosmology~\cite{Jamil:2011ptc,Shabani:2013djy,Shabani:2014xvi,Asghari:2024obf} and in the context of wormhole solutions~\cite{Azizi:2012yv,Zubair:2016cde,Moraes:2016akv,Moraes:2017mir,Elizalde:2018frj}.
These alternative frameworks not only extend the scope of GR but also offer promising pathways to addressing some of the universe’s most profound and unresolved questions.

Additionally, functions of the Gauss-Bonnet invariant, $G \equiv R^2 - 4R_{\mu\nu}R^{\mu\nu} + R_{\mu\nu\alpha\beta}R^{\mu\nu\alpha\beta}$,
can be incorporated into the gravitational action, leading to the $f(G)$ theories~\cite{Nojiri:2005jg,Bamba:2009gq,Nojiri:2010oco,Rodrigues:2012qu,Houndjo:2013us}, as well as extensions such as $f(G, R)$~\cite{Cognola:2006eg} and $f(G, T)$~\cite{Sharif:2016xjv,FarasatShamir:2017mlg,Bhatti:2017fov,Shamir:2018qhq} gravities.
Another intriguing gravitational formulation introduces the torsion tensor to describe gravitational interactions, known as teleparallel theory. An extension of this approach, which has greatly enhanced our understanding of cosmological phenomena, is the $f(\mathcal{T})$ theory~\cite{Bengochea:2008gz,Cai:2015emx}, where $\mathcal{T}$ denotes the torsion scalar. Notable generalizations include $f(\mathcal{T}, T)$~\cite{Kiani:2013pba,Harko:2014aja}, where $T$ represents the trace of the energy-momentum tensor, $f(\mathcal{T}, T_G)$~\cite{Kofinas:2014owa}, and $f(\mathcal{T}, \mathcal{B})$~\cite{Bahamonde:2015zma}, where $\mathcal{B}$ is the boundary term. For a detailed review of these propositions and their applications in various contexts, we recommend the following studies~\cite{Nunes:2016qyp,Paliathanasis:2016vsw,Bamba:2016gbu,Ahmed:2016cuy,Nunes:2018xbm,Cai:2018rzd,Ferraro:2018tpu,Nashed:2018cth,Nassur:2015zba,Junior:2015bva,Saez-Gomez:2016wxb,Pace:2017aon,Ghosh:2020rau,Kofinas:2014aka,Kofinas:2014daa,Bahamonde:2021srr,Bahamonde:2016cul,Caruana:2020szx,Escamilla-Rivera:2019ulu,Harko:2014sja}.
A separate class of theories is based on the non-metricity tensor, forming the foundation of symmetric teleparallel theory. Its extension, known as $f(\mathcal{Q})$ theory~\cite{BeltranJimenez:2017tkd,BeltranJimenez:2019tme}, employs $\mathcal{Q}$, the non-metricity scalar. Generalizations such as $f(\mathcal{Q}, T)$~\cite{Xu:2019sbp} and $f(\mathcal{Q}, \mathcal{B})$~\cite{Capozziello:2023vne,De:2023xua}, where $\mathcal{B}$ denotes the boundary term, have also been developed.
These frameworks have inspired a variety of compelling applications, including black hole solutions, which can be found in the following references~\cite{Anagnostopoulos:2021ydo,Khyllep:2021pcu,Frusciante:2021sio,Lin:2021uqa,Barros:2020bgg,DAmbrosio:2021zpm,Xu:2020yeg,Arora:2020tuk,Arora:2020iva,Arora:2020met,Agrawal:2021rur,Godani:2021mld,Arora:2021jik,Yang:2021fjy,Najera:2021afa,Shiravand:2022ccb,Heisenberg:2023lru,Pradhan:2023oqo,Junior:2023qaq,Junior:2024xmm}.

Beyond the previous discussion, another fundamental limitation of GR arises from one of its most well-known predictions. Karl Schwarzschild formulated a model for a compact object described solely by its mass. However, his solution features a central singularity, which is a region where the laws of physics break down, leading to infinite curvature and the loss of predictive power. Such singularities pose a major theoretical challenge, as they indicate an incomplete description of gravitational phenomena at extreme conditions.
To circumvent these regions where spacetime curvature diverges, various extensions of GR have been proposed, aiming to provide a more complete and physically meaningful framework. One promising approach involves coupling Einstein’s equations with electromagnetism. The resulting solutions, known as regular black holes (RBHs), prevent the formation of a central Schwarzschild singularity by modifying the structure of spacetime at small scales. The first RBH model was introduced by Bardeen \cite{Bardeen}, but it was only with the work of Ayón-Beato and García that this model was derived as an exact solution of GR, provided that the matter source is described by nonlinear electrodynamics (NLED) \cite{Ayon-Beato:2000mjt}.
NLED extends classical Maxwell electrodynamics by incorporating nonlinear effects, which become significant in strong electromagnetic fields, thereby offering a natural mechanism to regularize singularities. Inspired by these developments, numerous RBH models have been proposed, incorporating NLED as a matter source within the mathematical framework of GR \cite{Bronnikov:2000vy,Dymnikova:2004zc,Balart:2014cga,Culetu:2014lca,Rodrigues:2018bdc}. In a cosmological context, NLED has also been explored within GR, revealing that the accelerated expansion of the universe can naturally emerge from the electromagnetic quantities involved in these solutions, providing an intriguing alternative to dark energy models \cite{Novello:2003kh,Novello:2006ng}.

Accordingly, the primary objective of this manuscript is to investigate black hole solutions within the framework of $f(R,T)$ gravity, incorporating NLED as the matter source \cite{Tangphati:2023xnw}. To achieve this, we adopt a model characterized by the functional form $f(R,T) = R + \beta T$, where $\beta$ is a constant. Notably, when $\beta = 0$, this model reduces to GR, thereby serving as a generalization of Einstein's theory.
In addition to the gravitational framework, we consider a Lagrangian for the NLED field expressed as a power-law form: $\mathcal{L} = f_0 + F + \alpha F^p$, where $F$ is the Maxwell invariant, $\alpha$ is a coupling constant, and $p$ represents the power-law index. The constant term $f_0$ ensures the correct asymptotic behavior. Importantly, setting $\alpha = 0$ recovers the linear Maxwell Lagrangian, aligning this formulation with standard electromagnetic theory.
With these assumptions in place, we derive the black hole metric function by solving the equations of motion. This involves determining explicit solutions for the metric function under specific values of the power-law index $p$ and analyzing the properties of these solutions. A critical aspect of our analysis is to verify the regularity of the black hole solutions by evaluating curvature invariants, such as the Kretschmann scalar, to identify and characterize singularities.

Our second approach involves independently determining the functional forms of $\mathcal{L}_F(r)$ and $\mathcal{L}_{FF}(r)$ by using the components of the equations of motion. To do so, we adopted the metric function obtained in the previous approach, which is parametrized by the constants $\alpha$ and $\beta$. This allowed us to derive the general form of the nonlinear electrodynamics Lagrangian, $\mathcal{L}_{\text{NLED}}(F)$, under the influence of these constants.
From this analysis, we observe that the Lagrangian exhibits inherent nonlinearity, which can be attributed to the presence of the constants $\alpha$ and $\beta$. In particular, if we set $\alpha = 0$, the Lagrangian retains its nonlinear nature due to the contribution from the constant $\beta$. On the other hand, if the terms involving $\beta$ are zero, the nonlinearity persists solely as a result of $\alpha \neq 0$. 
While it is theoretically possible to recover a linear Lagrangian under specific conditions, the key takeaway is that even in the linear case, the Lagrangian remains within the framework of $f(R,T)$ gravity, as opposed to reducing to GR. This distinction is crucial, as it highlights the broader scope of our model in the context of modified theories of gravity.
Finally, we also use observational data from the Event Horizon Telescope (EHT) to estimate the shadow of the black hole predicted by our model. By comparing the theoretical shadow with the observed shadow of Sagittarius A* (Sgr A*), we were able to place constraints on the parameter $\beta$, offering a direct connection between our theoretical framework and astrophysical observations.

This work is organized as follows.  In Sec. \ref{sec2}, we present the field equations of $f(R,T)$ gravity coupled with nonlinear electrodynamics, define the class of geometries under consideration, and derive the field equations required to reconstruct the Lagrangian of the electromagnetic field in the context of a magnetic charge. In Sec. \ref{sec3}, we obtain several exact solutions of the field equations of the nonlinear electrodynamics in vacuum,  which generalize the known geometries of black holes, and we analyze their horizon properties. In Sec. \ref{rsh}, we study the shadow of the obtained black hole solutions, and we constrain the parameter $\beta$ of the model based on the data captured by the EHT for Sagittarius A*. Finally, in Sec. \ref{sec:concl}, we summarize and discuss our main results.

\section{Field equations of $f(R,T)$ gravity coupled to NLED}\label{sec2}

\subsection{Action and field equations}

In this work, we explore solutions for $f(R,T)$ gravity coupled with NLED, where the action is given by:
\begin{align}
S=\int\sqrt{-g}\, d^{4}x\left[f\left(R,T\right)+2\kappa^{2}{\cal L}_{\rm NLED}(F)\Big)\right]\,,
	\label{action}
\end{align}
where $g$ is the determinant of the metric $g_{\mu\nu}$, $\kappa^2=8\pi$, $R$ is the Ricci scalar and $T$ represents the trace of the energy-momentum tensor. The factor ${\cal L}_{\rm NLED}(F)$ is the nonlinear electromagnetic (NLED) Lagrangian density that depends on the electromagnetic scalar, $F=\frac{1}{4}F^{\mu\nu}F_{\mu\nu}$, where $F_{\mu \nu} = \partial_\mu A_\nu -\partial_\nu A_\mu$ is the antisymmetric Maxwell-Faraday tensor and ${A_\mu}$ is the electromagnetic vector potential.

Varying the action \eqref{action} with respect to $A_{\mu}$, we obtain the following equation of motion
\begin{eqnarray}
	\nabla_{\mu}\left[\left(2f_{T}\left(R,T\right){\cal L}_{FF}F-\kappa^{2}{\cal L}_{F}\right)F^{\mu\alpha}\right] &=&
	\nonumber
	\\ \hspace{-0,3cm}
	\frac{1}{\sqrt{-g}}\partial_{\mu}\left[\sqrt{-g}\left(2f_{T}\left(R,T\right){\cal L}_{FF}F-\kappa^{2}{\cal L}_{F}\right)F^{\mu\alpha}\right]	&=& 0\,,\label{solA}
\end{eqnarray}
where we have denoted ${\cal L}_F=\partial {\cal L} _{\rm NLED}(F)/\partial F$.
\par
On the other hand, the gravitational field equations are derived by varying the action \eqref{action} with respect to the metric tensor, which leads to \cite{Harko:2011kv}
\begin{eqnarray}
f_{R}R_{\mu\nu}-\frac{1}{2}fg_{\mu\nu}+\left(g_{\mu\nu}\square-\nabla_{\mu}\nabla_{\nu}\right)f_{R}
	\nonumber \\
=T_{\mu\nu}-f_{T}\left(T_{\mu\nu}+\Theta_{\mu\nu}\right)\,,
	\label{EqM}
\end{eqnarray}
where $f_{R}=\partial{f(R,T)}/\partial R$, $f_{T}=\partial{f(R,T)}/\partial T$ and $\square=g^{\mu\nu}\nabla_{\mu}\nabla_{\nu}$ is the d’Alembertian operator. 
In addition, the contributions of the matter tensors $T_{\mu\nu}$ and $\Theta_{\mu\nu}$ are defined as \cite{Harko:2011kv}
\begin{equation}
    T_{\mu\nu}=-\frac{2}{\sqrt{-g}}\frac{\delta {\cal L}_{\rm mat}\sqrt{-g}}{\delta g^{\mu\nu}} \,,
\end{equation}
and
\begin{equation}
   \Theta_{\mu\nu}=g^{\sigma\rho}\frac{\delta T_{\sigma\rho}}{\delta g^{\mu\nu}}\,,
\end{equation}
respectively, where ${\cal L}_{\rm mat}$ denotes the matter Lagrangian desnity.

In this work, our interest lies in the development of solutions where the matter Lagrangian of the energy-momentum tensor ${\cal L}_{\rm mat}$ is described by nonlinear electrodynamics. Therefore, for our purposes, the explicit form of these quantities in NLED is
\begin{align}
        \overset{F}{T}{}_{\mu\nu}=g_{\mu\nu}{\cal L}_{\rm NLED}(F)-{\cal L}_{F}F_{\mu\rho}F_{\nu}^{\phantom{\mu}\rho},
\end{align}
and
\begin{align}
   \Theta_{\mu\nu}=&-g_{\mu\nu}{\cal L}_{{\rm NLED}}(F)+F_{\mu\rho}F_{\nu}^{\phantom{\mu}\rho}
   \times
   \nonumber
   \\
   &
   \times\left(L_{F}\left(F\right)-\frac{1}{2}F_{\alpha\beta}F^{\alpha\beta}L_{FF}\left(F\right)\right).
\end{align}

\subsection{Field equations in static spherical symmetry}

To obtain solutions, we consider the following static and spherically symmetric metric in this article:
\begin{equation}
ds^2=A(r)dt^2-\frac{dr^2}{B(r)}-r^2 \left( d\theta^{2}+\sin^{2}\left(\theta\right)d\phi^{2} \right),\label{m}
\end{equation}
where $A(r)$ and $B(r)$  are functions of the radial coordinate $r$, and are independent of time.

Furthermore, the components of the $F_{\mu\nu}$ tensor that we consider in our solutions are described exclusively by the magnetic charge $q$:
\begin{align}
        F_{23}=-F_{32}&=q \sin\theta \,,\label{mag}
\end{align}
where the electromagnetic scalar $F$ now takes the form
\begin{equation}
    F=\frac{q^{2}}{2r^{4}}.\label{F1}
\end{equation}
Note that component \eqref{mag} satisfies the modified Maxwell equation \eqref{solA}.
\par
The following relationships will be important to check the coherence of the solutions that we find below
\begin{equation}
    {\cal L}_F=\frac{\partial {\cal L} _{\rm NLED}}{\partial r} \bigg(\frac{\partial F}{\partial r}\bigg)^{-1}\label{RC}
\end{equation}
and
\begin{equation}
   {\cal L}_{FF}=\frac{\partial{\cal L}_{F}}{\partial r}\bigg(\frac{\partial F}{\partial r}\bigg)^{-1}.\label{RC2}
\end{equation}

With all these tools at our disposal, the components of the equations of motion \eqref{EqM} with mixed indices are given by:
\begin{eqnarray}
		&&\frac{Bf_{R}A''}{2A}+\frac{f_{R}A'B'}{4A}-
		\frac{Bf_{R}A'^{2}}{4A^{2}}+\frac{Bf_{R}A'}{rA}
		-\frac{1}{2}B'f_{R}'
			\nonumber \\
			&&\hspace{2cm}
		-Bf_{R}''-\frac{f}{2}
		-\frac{2Bf_{R}'}{r}=2{\cal L}_{\rm NLED}\,,\label{EqF00}
\end{eqnarray}
\begin{eqnarray}
	&&\frac{Bf_{R}A''}{2A}+\frac{f_{R}A'B'}{4A}
	-\frac{BA'f_{R}'}{2A}-\frac{Bf_{R}A'^{2}}{4A^{2}}
	\nonumber
	\\
	&&\hspace{2cm} +\frac{f_{R}B'}{r}-\frac{2Bf_{R}'}{r}-\frac{f}{2}=2{\cal L}_{\rm NLED}\,,\label{EqF11}
\end{eqnarray}
\begin{eqnarray}
	&&-\frac{BA'f_{R}'}{2A}+\frac{Bf_{R}A'}{2rA}
	-\frac{1}{2}B'f_{R}'+\frac{f_{R}B'}{2r}-Bf_{R}''
	-\frac{Bf_{R}'}{r}
	\nonumber\\
	&&\hspace{2cm}
	+\frac{Bf_{R}}{r^{2}}-\frac{f}{2}
	-\frac{f_{R}}{r^{2}}=2{\cal L}_{{\rm NLED}}
			\nonumber\\
	&&\hspace{3cm}
	+\frac{2q^{4}f_{T}{\cal L}_{FF}}{r^{8}}
	-\frac{2q^{2}{\cal L}_{F}}{r^{4}}\,,\label{EqF22}
\end{eqnarray}
where all the quantities are functions of the radial coordinate $r$, and the prime denotes a derivative with respect to $r$.

To analyze the properties of the metric functions that we present later and to determine the presence of horizons, we use the following condition:
\begin{equation}
A(r_{H})=0.\label{rH}
\end{equation}
where the radius $r_H$ denotes the presence of an horizon.
In addition, a second condition allows us to determine the number of horizons in relation to the metric function:
\begin{equation}
\frac{d A(r_{H})}{dr}\bigg|_{r=r_H}=0,
\label{der_a}
\end{equation}

We use these two relations to identify degenerate horizons and determine the critical values of the model parameters that we will obtain in the following sections.

\section{Solution of $f(R,T)$ gravity coupled to NLED}\label{sec3}

In this section, we derive solutions within the framework of $f(R,T)$ gravity, focusing on the specific case where the functional form of $f(R,T)$ is linear and includes a constant, $\beta$, coupled to the trace of the energy-momentum tensor. Since GR is the theory that most closely aligns with cosmological data and has been extensively validated through solar system tests, we anticipate any deviation introduced by $f(R,T)$ gravity to be minimal. Guided by this expectation, we adopt the following explicit functional form for $f(R,T)$:
\begin{equation}
 f\left(R,T\right) = R + \beta\, T.\label{function}
\end{equation}

As detailed in Section \ref{top1}, our initial approach involves deriving solutions by proposing a generalization of the Lagrangian for NLED. In this framework, we assume that the NLED Lagrangian follows an arbitrary power-law form, which serves as the basis for our investigation.

In Section \ref{top2}, we proceed to determine the general expressions for the first and second derivatives of the NLED Lagrangian by using the equations of motion. These derivatives enable us to reconstruct the analytical form of the NLED Lagrangian as a function of the electromagnetic scalar $F$, providing further insight into the interplay between the Lagrangian structure and the underlying field dynamics.

\subsection{Solutions for power law forms of NLED Lagrangian}\label{top1}

In this solution, we consider a generalized form of the Lagrangian for nonlinear electrodynamics, characterized explicitly by a power-law formulation. This approach allows us to explore the impact of the arbitrary power-law exponent on the resulting field dynamics and spacetime geometry, providing a versatile framework for analyzing the interplay between NLED and $f(R,T)$ gravity. Thus, consider the following expression of the Lagrangian of the nonlinear electromagnetic field
\begin{align}
  &  {\cal L}_{{\rm NLED}}\left(F\right)=f_{0}+F+\alpha F^{p},	\label{L}
\end{align}
giving
\begin{align}
{\cal L}_{F}\left(F\right)=1+\alpha \,p\, F^{p-1},	\label{LF}
\end{align}
and
\begin{align}	
{\cal L}_{FF}\left(F\right)=\alpha\left(p-1\right)F^{p-2},\label{LFF}
\end{align}
respectively.

By considering the form of the function \eqref{function}, along with the expressions (\ref{L})--(\ref{LFF}) in the equations of motion, and assuming the symmetry $B(r) = A(r)$, we derive the following metric function from the components of the field equations (\ref{EqF00}-\ref{EqF22})
\begin{eqnarray}
A(r)&=& 1-\frac{2M}{r}+\frac{q^{2}}{r^{2}} -\frac{2}{3}(2\beta+1)f_{0}r^{2}
\nonumber \\
&&
+\frac{2^{1-p}}{3-4p}\alpha\,\left[2\beta\left(p-1\right)-1\right]q^{2p}r^{2-4p}\,.\label{A}
\end{eqnarray}
Indeed, by identifying an effective cosmological constant as 
\begin{equation}
    \Lambda_{\rm eff} = 2\,(2\beta + 1)f_{0},\label{Lambda}
\end{equation}
we can rewrite the metric function~\eqref{A} as follows:
\begin{align}
A(r)=& 1-\frac{2M}{r}+\frac{q^{2}}{r^{2}} -\frac{\Lambda_{\rm eff}}{3}r^{2}
\nonumber
\\
&
+\frac{2^{1-p}}{3-4p}\alpha\,\left[2\beta\left(p-1\right)-1\right]q^{2p}r^{2-4p}\,.\label{A1}
\end{align}

Significant implications arise from the constraints that can be imposed on the constants appearing in the metric function, Eq.~\eqref{A1}. 
First, by setting $\alpha = 0$, the metric function reduces to a Reissner-Nordström-AdS type solution, reflecting the expected behavior when the nonlinear terms in the electrodynamics are eliminated. Furthermore, the Schwarzschild metric is recovered under the assumptions of zero charge ($q = 0$), a vanishing effective cosmological constant ($\Lambda_{\text{eff}} = 0$), and $\alpha = 0$, indicating the consistency of the solution with GR under these specific conditions.
It is also noteworthy that if we impose $\beta = \frac{1}{2(p-1)}$, the last term in Eq.~\eqref{A1} vanishes. Despite this simplification, the resulting solution still belongs to the framework of $f(R,T)$ gravity. This retention of the $f(R,T)$ nature is due to the inherent nonlinearity present in the Lagrangian.

The electromagnetic quantities, after substituting Eq. \eqref{F1} into Eqs. \eqref{L}--\eqref{LFF}, are now described in terms of r as:
\begin{equation}
    {\cal L}_{\text{NLED}}(r)	=f_{0}+\frac{q^{2}}{2r^{4}}+\alpha2^{-p}\left(\frac{q^{2}}{r^{4}}\right)^{p}\,,\label{Lxr}
\end{equation}
\begin{equation}
{\cal L}_{F}(r)	= 1 + \alpha2^{1-p}p\left(\frac{q^{2}}{r^{4}}\right)^{p-1}\,,
\label{LFxr}
\end{equation}
\begin{equation}
{\cal L}_{FF}(r)	=\alpha2^{2-p}\left(p-1\right)p\left(\frac{q^{2}}{r^{4}}\right)^{p-2} \,.
\label{LFFxr}
\end{equation}

Next, we proceed to explore specific solutions by assigning distinct powers to the metric function given in Eq.~\eqref{A1}. Our focus will be on the cases where the power takes the values $p = 2$, $p = 4$, and $p = 6$.
Among these, we will provide a detailed analysis of the solution corresponding to $p = 6$, as this case serves as a representative example that can be readily extended to other powers. The methodology and insights obtained from this detailed exploration will facilitate the development of analogous solutions for the remaining powers.

Furthermore, as we shall demonstrate in more detail later, even when deriving the Lagrangian in terms of the electromagnetic scalar $F$ within a more general framework, the use of the metric function given in Eq.~\eqref{A1} ensures that the resulting theory retains the characteristics of nonlinear electrodynamics (NLED), even in the absence of the constant $\alpha$ introduced in the Lagrangian \eqref{L}. 
This persistence of nonlinearity arises because the Lagrangian expressed in terms of $F$ contains terms reflecting the contributions of both $\alpha$ and $\beta$. Here, $\beta$ is the constant associated with the $f(R,T)$ function defined in Eq.~\eqref{function}. For the Lagrangian to achieve a strictly linear form, all terms involving these constants must vanish. 
Nevertheless, even under these restrictive conditions—representing the linear case—the underlying framework remains within the $f(R,T)$ theory, highlighting its robustness and the inherent modifications it introduces to gravitational dynamics compared to standard General Relativity.

\subsubsection{Solution with $p=2$}

For this solution, we consider $p = 2$ in the metric function given by \eqref{A1}, resulting in the following form:
\begin{align}
A\left(r\right)=1-\frac{2M}{r}+\frac{q^{2}}{r^{2}}-\frac{\Lambda_{\rm eff} r^{2}}{3}-\frac{\alpha\left(2\beta-1\right)q^{4}}{10r^{6}}
.\label{Ap2}
\end{align}

The Lagrangian in terms of the radial coordinate, Eq. \eqref{Lxr}, for this case becomes
\begin{align}
  {\cal L}_{{\rm NLED}}\left(r\right)=f_{0}+\frac{q^{2}}{2r^{4}}+\frac{\alpha q^{4}}{4r^{8}}.\label{Lxr_p2}
\end{align}
In principle, we can rewrite $r(F)$ from Eq. \eqref{F1}. Thus, for this model, we obtain the Lagrangian in terms of $F$, described by
\begin{align}
  {\cal L}_{{\rm NLED}}\left(F\right)=\frac{\rm \Lambda_{eff}}{4\beta+2}+F+\alpha F^{2}.\label{Lxf_p2}
\end{align}
For small values of $F$, we verify that the lagrangian behaves as follows
\begin{equation}
    {\cal L}_{{\rm NLED}}\left(F\right)\sim\frac{\Lambda_{\rm eff}}{4\beta+2}+F
\end{equation}
while for large values of $F$, the Lagrangian behaves approximately as follows
\begin{equation}
    {\cal L}_{{\rm NLED}}\left(F\right)\sim\frac{\Lambda_{\rm eff}}{4\beta+2}+F+\alpha F^2
\end{equation}
Note that the weak field limit is only recovered for small values of $F$.

We also calculated the Kretschmann scalar for this solution, with its explicit form given by
\begin{eqnarray}
    K(r)&=&\Big[-40\alpha(2\beta-1)q^{4}r^{4}\left(-84Mr+114q^{2}-5\Lambda_{\rm eff} r^{4}\right)
    \nonumber
\\
&&
+200r^{8}\left(18M^{2}r^{2}-36Mq^{2}r+21q^{4}+\Lambda_{\rm eff}^{2}r^{8}\right)
\nonumber
\\
&&
+1434\alpha^{2}(1-2\beta)^{2}q^{8}\Big]
\Big/ 75r^{16}.\label{k2}
\end{eqnarray}
We verify  that the Kretschmann scalar~\eqref{k2}, diverges for small values of $r$, while at asymptotic infinity, i.e., for large values of $r$, it behaves as $8 \Lambda_{\rm eff}^2 / 3$.

\subsubsection{Solution with $p=4$}

We now consider the solution for $p=4$. With this imposition, the metric function \eqref{A1} takes the following form:
\begin{align}
A(r)=1-\frac{2M}{r}+\frac{q^{2}}{r^{2}}-\frac{\Lambda_{eff}}{3}r^{2}-\frac{\alpha\,\left[6\beta-1\right]q^{8}}{104r^{14}}
.\label{Ap4}
\end{align}
Thus, the Lagrangian in terms of the radial coordinate is 
\begin{align}
  {\cal L}_{{\rm NLED}}\left(r\right)=f_{0}+\frac{q^{2}}{2r^{4}}+\frac{\alpha q^{8}}{16r^{16}}.\label{Lxr_p4}
\end{align}

By writing $r(F)$, we find that the Lagrangian is now described as follows:
\begin{align}
  {\cal L}_{{\rm NLED}}\left(F\right)=\frac{\Lambda_{\rm eff}}{4\beta+2}+F+\alpha F^{4}.\label{Lxf_p4}
\end{align}
For small values of $F$ we observe that the Lagrangian behaves as follows
\begin{equation}
{\cal L}_{{\rm NLED}}\left(F\right)\sim\frac{\Lambda_{\rm eff}}{4\beta+2}+F \,,
\end{equation}
while for large values of $F$ the Lagrangian behaves approximately as
\begin{equation}
{\cal L}_{{\rm NLED}}\left(F\right)\sim\frac{\Lambda_{\rm eff}}{4\beta+2}+F+\alpha F^2.
\end{equation}
As in the previous case, the linear case is recovered only for small values of $F$.

The Kretschmann scalar for this configuration is
\begin{align}
    &K(r)=-\frac{2\alpha(6\beta-1)q^{8}\left(-120Mr+172q^{2}-13\Lambda_{\rm eff} r^{4}\right)}{13r^{20}}
    \nonumber
\\
&
+\frac{5611\alpha^{2}(1-6\beta)^{2}q^{16}}{1352r^{32}}
+\frac{8\left(6M^{2}r^{2}-12Mq^{2}r+7q^{4}\right)}{r^{8}}
\nonumber
\\
&
+\frac{8\Lambda_{\rm eff}^{2}}{3}.\label{k4}
\end{align}

The Kretschmann scalar, as described by Eq.~\eqref{k4}, diverges for small values of $r$ and approaches the value $8 \Lambda_{\rm eff}^2 / 3$ for large values of $r$, as in the previous example.

\subsubsection{Solution with $p=6$}

Finally, we consider $p=6$. In this case, the metric function \eqref{A1} takes the following form:
\begin{align}
A(r)=1-\frac{2M}{r}+\frac{q^{2}}{r^{2}}-\frac{\Lambda_{\rm eff}}{3}r^{2}-\frac{\alpha(10\beta-1)q^{12}}{672r^{22}}.\label{Ap6}
\end{align}
which provides the following Lagrangian
\begin{align}
  {\cal L}_{{\rm NLED}}\left(r\right)=f_{0}+\frac{q^{2}}{2r^{4}}+\frac{\alpha q^{12}}{16r^{24}}.\label{Lxr_p6}
\end{align}

If we write $r(F)$, we find that the Lagrangian is now described by:
\begin{align}
{\cal L}_{{\rm NLED}}\left(F\right)=\frac{\Lambda_{\rm eff}}{4\beta+2}+F+\alpha F^{6}.\label{Lxf_p6}
\end{align}
For $F \ll 1$ we have
\begin{equation}
{\cal L}_{{\rm NLED}}\left(F\right)\sim\frac{\Lambda_{\rm eff}}{4\beta+2}+F \,,
\end{equation}
while for large values of $F$ the Lagrangian behaves approximately as follows:
\begin{equation}
{\cal L}_{{\rm NLED}}\left(F\right)\sim\frac{\Lambda_{\rm eff}}{4\beta+2}+F+\alpha F^6.
\end{equation}
Again, the linear case is recieveed if $F$ is very small.

Now the Kretschmann scalar is given as
\begin{align}
    &K(r)=-\frac{\alpha(10\beta-1)q^{12}\left(-276Mr+402q^{2}-35\Lambda_{\rm eff} r^{4}\right)}{42r^{28}}
    \nonumber
\\
&
+\frac{3583\alpha^{2}(1-10\beta)^{2}q^{24}}{6272r^{48}}
+\frac{8\left(6M^{2}r^{2}-12Mq^{2}r+7q^{4}\right)}{r^{8}}
\nonumber
\\
&
+\frac{8\Lambda_{\rm eff}^{2}}{3}.\label{k6}
\end{align}
Thus, as in the previous cases, the Kretschmann scalar, as presented in Eq.~\eqref{k6}, also diverges for small values of $r$ and approaches $8 \Lambda_{\rm eff}^2 / 3$ for large values of $r$.

\begin{figure}[th!]
\includegraphics[scale=0.57]{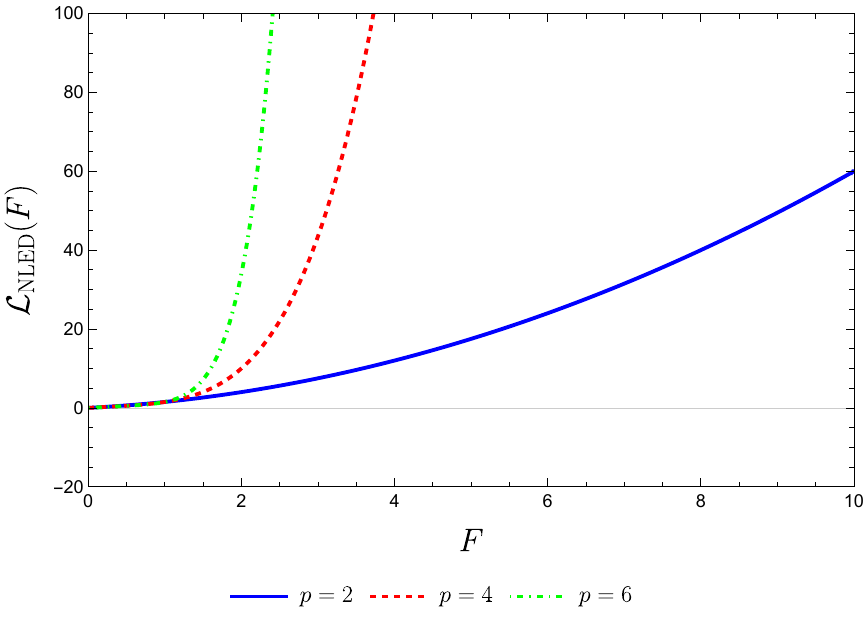}
\caption{The NLED Lagrangian ${\cal L}(F)$, for the Eqs. \eqref{Lxf_p2}, \eqref{Lxf_p4} and \eqref{Lxf_p6}. We have used the values of the constants as follows $f_0=0.01$, $\alpha=0.5$, $\beta=0.25$.}
\label{fig_LxF}
\end{figure}

In Fig. \ref{fig_LxF} we illustrate the behavior of the Lagrangian densities of NLED as a function of $F$, described by Eqs. \eqref{Lxf_p2}, \eqref{Lxf_p4} and \eqref{Lxf_p6}, i.e., for the powers $p=2$, $p=4$ and $p=6$, respectively.
%

\subsubsection{Horizons}

Next, we present the horizon graphs, focusing exclusively on the metric function with the power $p = 6$, as the graphical behavior for the other powers is analogous. 

Below, we illustrate the event horizon representations obtained by solving Eqs.~\eqref{rH} and \eqref{der_a} simultaneously, applied to the metric function given in Eq.~\eqref{Ap6}. This approach allows us to determine key parameters such as the critical electric charge $q_{\text{ec}}$ and the critical mass $M_{\text{ec}}$. 

\begin{figure}[t!]
\includegraphics[scale=0.57]{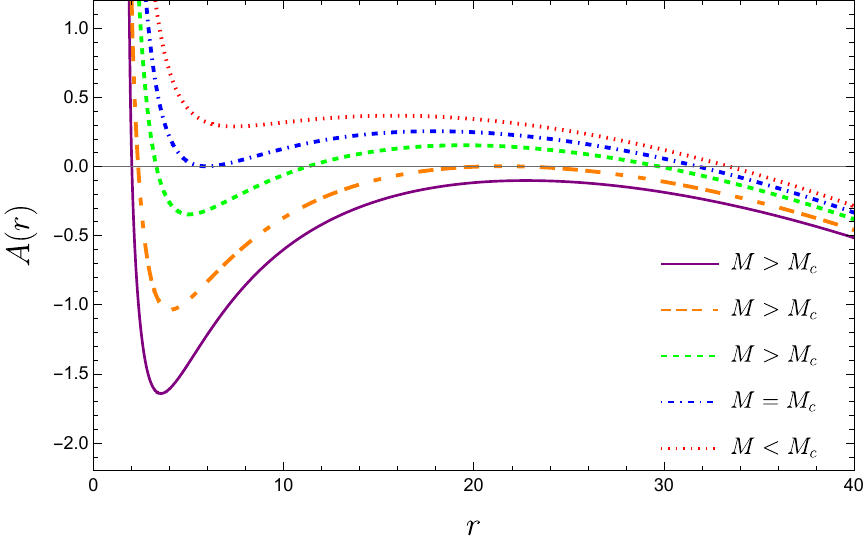}
\caption{The plot of the metric function  $A(r)$, described by Eq.~\eqref{Ap6}, for the values $\alpha=0.5$, $\beta =0.002$, $f_0 =0.001$  and $q=5.75$. } 
\label{fig_ap6_Mc}
\end{figure}

For instance, we determined the critical mass to be \(M_c = 5.682\) by solving these equations under the following parameter values: \(\alpha = 0.5\), \(\beta = 0.002\), \(f_0 = 0.001\), and \(q = 5.75\). Figure~\ref{fig_ap6_Mc} illustrates the behavior of the metric function \eqref{Ap6} as a function of the radial coordinate \(r\) for three distinct mass scenarios: \(M > M_c\), \(M = M_c\), and \(M < M_c\).

(i) $M > M_c$:  
Three different scenarios emerge depending on the mass value. For larger values of $M > M_c$, three horizons are observed, as depicted by the green curve; for slightly smaller values of $M > M_c$, two horizons are present, as depicted by the orange curve; Ffor $M > M_c$ near $M_c$, only one horizon appears, as represented by the purple curve.

(ii) $M = M_c$:   
In this scenario, two horizons are observed. One of these is a degenerate horizon, while the outermost remains the cosmological horizon.

(iii) $M < M_c$: For this case, only the cosmological horizon exists, as indicated by the absence of any additional inner horizons.

These results provide a clear depiction of the dependence of the horizon structure on the critical mass $M_c$ and demonstrate the transition between scenarios as $M$ varies relative to $M_c$.\\

\begin{figure}[t!]
\includegraphics[scale=0.57]{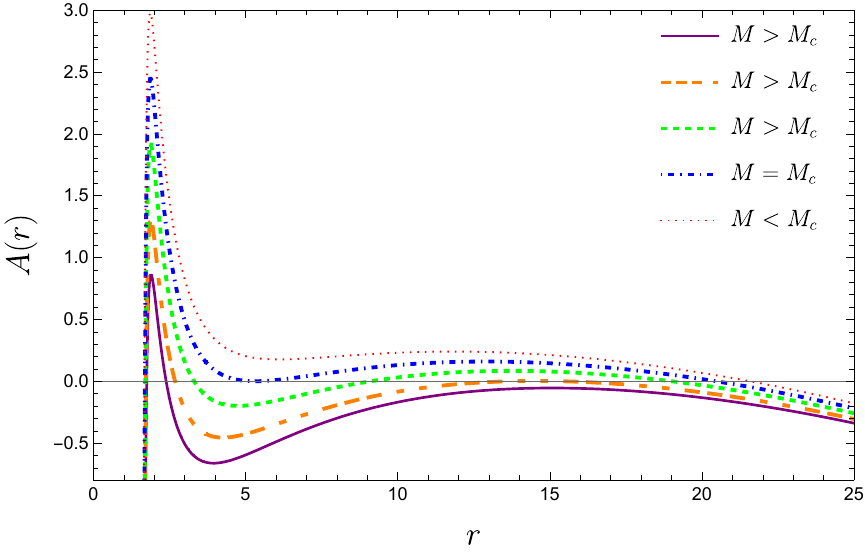}
\caption{ The plot of the metric function  $A(r)$, described by Eq.~\eqref{Ap6}, for the values $\alpha=-1.5$, $\beta =0.02$, $f_0 =0.001$  and $q=5.75$.} 
\label{fig_ap6_Mc2}
\end{figure}

We revisit the simultaneous solution of Eqs.~\eqref{rH} and \eqref{der_a}, applying the metric function \eqref{Ap6} to derive an additional solution for the critical mass. In this analysis, we obtained a critical mass value of $M_c = 5.680$, considering the constants $\alpha = -1.5$, $\beta = 0.02$, $f_0 = 0.001$, and $q = 5.75$. The behavior of the metric function \eqref{Ap6} with respect to the radial coordinate $r$ is illustrated in Fig.~\ref{fig_ap6_Mc2}, highlighting three distinct mass scenarios: $M > M_c$, $M = M_c$, and $M < M_c$.

(i) $M > M_c$:
For this mass regime, three scenarios are observed:
For sufficiently large \(M > M_c\), up to four horizons are present, as depicted by the green curve; for slightly smaller $M > M_c$, three horizons are observed, as represented by the orange curve; near $M_c$, $M > M_c$ exhibits two horizons, as depicted by the purple curve.

(ii) $M = M_c$:
At the critical mass $M_c = 5.680$, the metric function exhibits three horizons: (a) an innermost internal horizon, (b) the event horizon, and, (c) the outer cosmological horizon.

(iii) $M < M_c$: For masses below the critical threshold ($M < M_c$), only two horizons are present, namely, (a) the innermost horizon, corresponding to the event horizon and (b) the outermost cosmological horizon.

In all cases, the metric function becomes negative for large values of $r$, consistent with the behavior expected in this model. This result reinforces the intricate relationship between the constants and the horizon structure, as well as the significance of the parameter choices in determining the physical properties of the black hole spacetime.\\

\begin{figure}[t!]
\includegraphics[scale=0.57]{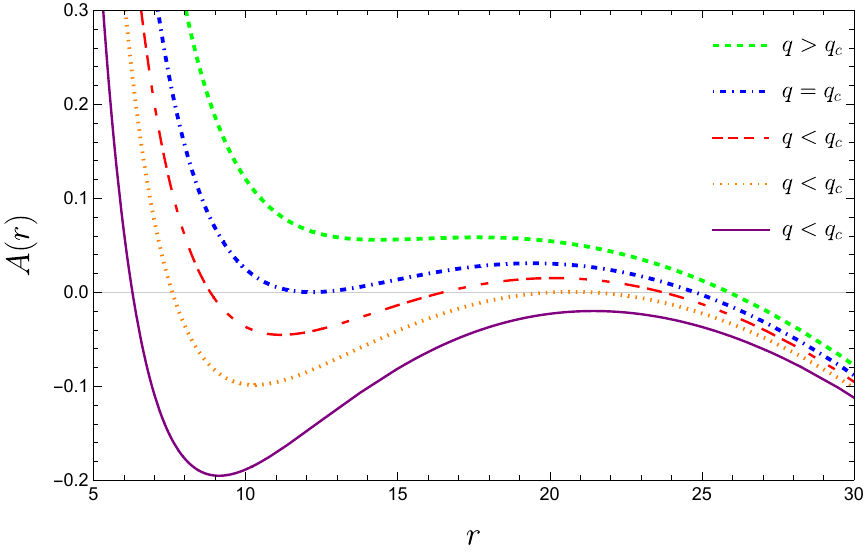}
\caption{  The plot of the metric function  $A(r)$, described by Eq.~\eqref{Ap6}, for the values $\alpha = 0.45$, $\beta = 0.0019$, $f_{0} = 0.0009$, and $M = 10.0$.} 
\label{fig_ap6_qc}
\end{figure}

Analogous to the determination of the critical mass, we now calculate the critical electric charge by solving Eqs.~\eqref{rH} and \eqref{der_a} simultaneously, utilizing the metric function given in Eq.~\eqref{Ap6}. Through this approach, we find the critical charge to be $q_{c} = 10.416$, with the constants chosen as $\alpha = 0.45$, $\beta = 0.0019$, $f_{0} = 0.0009$, and $M = 10.0$. The behavior of the metric function \eqref{Ap6} as a function of the radial coordinate $r$ is depicted in Fig.~\ref{fig_ap6_qc}, showcasing three scenarios of electric charge: $q_e > q_{ec}$, $q_e = q_{ec}$, and $q_e < q_{ec}$.

(i) $q_e > q_{c}$: When the electric charge exceeds the critical value ($q_e > q_{c}$), the metric function reveals the presence of a single horizon. 

(ii) $q_e = q_{c}$:  
At the critical charge ($q_e = q_{c}$), the solution exhibits two event horizons. 

(iii) $q_e < q_{c}$:  
For electric charges below the critical threshold ($q_e < q_{c}$), three distinct scenarios arise: For sufficiently small $q_e < q_{c}$, up to three horizons are observed, as depicted by the red curve; for intermediate values of $q_e < q_{c}$, the solution shows two horizons, as represented by the orange curve; for $q_e < q_{c}$ near the threshold, only a single horizon remains, as depicted by the purple curve.

These results underscore the dependence of the horizon structure on the electric charge and demonstrate the intricate interplay between the parameters $\alpha$, $\beta$, and $f_0$ within the framework of the $f(R,T)$ theory and NLED.

\subsection{Lagrangian reconstruction} \label{top2}

To further develop the solutions within this section, we revisit the function given in Eq.~\eqref{function}. The strategy employed here involves determining the general forms of the first derivative of the Lagrangian, ${\cal L}_{F}(r)$, and the second derivative of the Lagrangian, ${\cal L}_{FF}(r)$. These quantities are derived directly from the components of the equations of motion, namely Eqs.~\eqref{EqF00} and \eqref{EqF22}. In this manner, we obtain the following expressions:
\begin{align}
   {\cal L}_{F}\left(r\right)=\frac{r^{2}\left[r\left[A'(r)+2{\cal L}_{{\rm NLED}}(r)\left(2\beta r+r\right)\right]+A(r)-1\right]}{2\beta q^{2}},\label{LF1}
\end{align}
and
\begin{align}
   {\cal L}_{FF}\left(r\right)=&r^6\Big\{r\Big[-\beta rA''(r)+2A'(r)+4L(r)\left(2\beta r+r\right)\Big],
   \nonumber
   \\
   &
+2(\beta+1)A(r)-2(\beta+1)\Big\}\big/\left(4\beta^{2}q^{4}\right),	\label{LFF1}
\end{align}
respectively. 
The derived quantities ${\cal L}_{F}(r)$ and ${\cal L}_{FF}(r)$ satisfy all components of the equations of motion as well as the consistency relation given in Eq.~\eqref{RC2}. Consequently, the expressions in Eqs.~\eqref{LF1} and \eqref{LFF1} are independent on the specific choice of the metric function, allowing flexibility in selecting any metric function that satisfies Eq.~\eqref{RC2}. 
To proceed with further calculations, we will employ the metric function derived in the solution presented in the previous section, as given by Eq.~\eqref{A1}.

It is important to highlight that we now have two consistency relations applicable to the $f(R,T)$ theory, namely Eqs.~\eqref{RC} and \eqref{RC2}. Initially, it is unclear whether these relations will yield distinct solutions when used individually to develop a specific solution. 
In the following discussion, we will analyze the results obtained by employing each of these consistency relations separately. First, we will present the outcomes derived using Eq.~\eqref{RC}, followed by the results obtained when using the second relation, Eq.~\eqref{RC2}.

\subsubsection{Strategy one}

Subsequently, by substituting the derivative of the Lagrangian density, as described in Eq.~\eqref{LF1}, into the consistency relation given by Eq.~\eqref{RC}, we derived an expression that enabled us to determine the Lagrangian density in the following form:
\begin{align}
  {\cal L}_{{\rm NLED}}(r)=&-r^{-\frac{2}{\beta}-4}\int\frac{r^{\frac{\beta+2}{\beta}}\left(rA'(r)+A(r)-1\right)}{\beta}\,dr
  \nonumber
   \\
   &
  +f_{1} r^{-\frac{2}{\beta}-4}
   .\label{L1}
\end{align}

Substituting the metric function from Eq.~\eqref{A} into the Lagrangian \eqref{L1} yields
\begin{equation}
    {\cal L}_{{\rm NLED}}\left(r\right)=f_{0}+\frac{q^{2}}{2r^{4}}+f_{1}r^{-\frac{2}{\beta}-4}+\alpha2^{-p}q^{2p}r^{-4p}.\label{lagrang}
\end{equation}
Then, inverting to find $r(F)$ from Eq.~\eqref{F1}, we rewrite the Lagrangian \eqref{lagrang} as:
\begin{equation}
    {\cal L}_{{\rm NLED}}\left(F\right)=f_{0}+F+\alpha F^{p}+2^{\frac{1}{2\beta}+1}f_{1}F^{\frac{1}{2\beta}+1}q^{-\frac{1}{\beta}-2}.\label{Lag_F}
\end{equation}
This represents the general form of the Lagrangian in terms of $F$ with arbitrary power. Note that by setting $f_{0} = f_{1} = \alpha = 0$, we recover the Maxwell Lagrangian. 
By imposing $f_1=0$, we recover the form of the Lagrangian given by \eqref{L}.

It is important to note that the nonlinearity of the Lagrangian \eqref{Lag_F} is not solely associated with the constant $\alpha$. Even if this constant is set to zero, i.e., by taking $\alpha = 0$ in \eqref{Lag_F}, the Lagrangian remains nonlinear due to the presence of another term involving $F$, where the constant $\beta$ appears as an exponent. In this scenario, the Lagrangian \eqref{Lag_F} can be expressed in the following form:
\begin{equation}
    {\cal L}_{{\rm NLED}}\left(F\right)=f_{0}+F+2^{\frac{1}{2\beta}+1}f_{1}F^{\frac{1}{2\beta}+1}q^{-\frac{1}{\beta}-2}.\label{Lag2_F}
\end{equation}
Recall that the constant $\beta$ arises from the proposed choice of a linear function in $f(R,T)$ gravity, where this constant is coupled to the trace of the energy-momentum tensor, as described in Eq.~\eqref{function}. Moreover, if we were to remove only the term involving $F$ with the exponent associated with $\beta$ in the Lagrangian \eqref{Lag_F}, the theory would still fall under the category of NLED. This is because, with $\alpha \neq 0$, the Lagrangian \eqref{Lag_F} retains a term with an arbitrary exponent. In this case, the Lagrangian would take the following form:
\begin{equation}
    {\cal L}_{{\rm NLED}}\left(F\right)=f_{0}+F+\alpha F^p.\label{Lag2_Fb}
\end{equation}
Therefore, in order to recover the weak field limit, we can consider in the Lagrangian \eqref{Lag_F} $\alpha=0$ and $f_1=0$.

In the case of the Lagrangian \eqref{Lag_F}, if we choose $\beta = -\frac{1}{2}$ and make an appropriate choice for the constant $f_1$, the Lagrangian \eqref{Lag_F} becomes similar to Eq.~\eqref{Lag2_F}. However, even under this condition, the theory remains within the framework of $f(R,T)$ gravity. On the other hand, if the constant $\beta$ tends toward infinity and $\alpha = 0$, the theory transitions to a linear case.

To explore the behavior of the theory as $\beta$ approaches zero, the constant $f_1$ must be carefully chosen such that this new term vanishes rapidly, effectively canceling out all terms containing $\beta$. Mathematically, this requires $f_1$ to depend on $\beta$ in a manner that ensures the cancellation. Alternatively, a simpler method to evaluate this limit is to directly set $\beta = 0$ in the components of the equations of motion, given by Eqs.~(\ref{EqF00})--(\ref{EqF22}).

\subsubsection{Strategy two}

In this second approach, we determine the Lagrangian in terms of the electromagnetic scalar, ${\cal L}_{{\rm NLED}}(F)$, by employing the second consistency relation, Eq.~\eqref{RC2}. The process is analogous to that described in the previous section, with the key difference being the initial use of Eq.~\eqref{RC2} instead of Eq.~\eqref{RC}. 
Specifically, by substituting the derivative of the Lagrangian density, Eq.~\eqref{LF1}, into Eq.~\eqref{RC2}, we obtain the same Lagrangian density as previously derived, which is described by Eq.~\eqref{L1}. 

Furthermore, when the metric function is specified as Eq.~\eqref{A}, we deduce the corresponding Lagrangian in terms of the radial coordinate $r$, as expressed in Eq.~\eqref{lagrang}.
The analytical form that we obtain is the same as that described by Eq.~\eqref{Lag_F}.
Thus, we observe that there is no distinction between using either relation \eqref{RC} or \eqref{RC2}, as both provide the same analytical solution for the Lagrangian ${\cal L}_{{\rm NLED}}(F)$.

\section{Effective metric and shadow radius}\label{rsh}

In this section, we will compare the estimated shadow radius for our model of $f(R,T)$ gravity, described by the power law in the Lagrangian \eqref{lagrang} with the specific function given by Eq.~\eqref{function}, to the size of the shadow radius of the supermassive black hole at the center of our galaxy, Sagittarius (Sgr.) A*, as observed by the Event Horizon Telescope (EHT). 
To calculate the black hole shadow for our model, we will employ the method developed by \cite{Perlick:2021aok}. This approach allows for a systematic evaluation of the shadow radius based on the characteristics of the spacetime geometry derived from our modified gravity framework.
The estimated shadow size limits for the black hole Sgr A*, as observed by the EHT, are summarized in Table~\ref{tab:my-table}. These values incorporate both theoretical and observational uncertainties as reported in \cite{Vagnozzi:2022moj}. The data provides a benchmark for assessing the compatibility of our model with the current observational constraints:
\begin{equation} \label{eq:1sigma}
    4.55 \lesssim r_{sh}/M \lesssim 5.22 \ ,
\end{equation}
at $1\sigma$ deviation, while
\begin{equation} \label{eq:2sigma}
    4.21 \lesssim r_{sh}/M \lesssim 5.56 \ ,
\end{equation}
at $2\sigma$ deviation.
With this, we will restrict the parameter $\beta$, which is present in our black hole model.

\subsection{Effective metric}

However, having obtained solutions to the equations of the $f(R,T)$ theory coupled with NLED, described by the line element \eqref{m}, the light rays now follow a geodesic defined by the effective metric. In this way, we present the correct formulation for the metric~\cite{Novello:1999pg}
\begin{equation}
     g_{\rm eff}^{\mu\nu}={\cal L}_{F}g^{\mu\nu}-{\cal L}_{FF}F_{\sigma}^{\phantom{\sigma}\mu}F^{\sigma\nu}.\label{g_efet}
 \end{equation}
 Therefore, we have two line elements, depending on whether we consider the electric or magnetic charge. Given that we are considering a magnetic charge, Eq.~\eqref{mag}, the line element for the effective metric is consequently given by
\begin{eqnarray}
    ds^{2}=\bar{A}dt^{2}-\bar{B}(r)dr^{2}-\bar{C}^{2}(r)\left(d\theta^{2}+\sin^{2}\theta d\phi^{2}\right).\label{ds_mag}
\end{eqnarray}
where the explicit forms of the metric functions are \cite{Toshmatov:2021fgm}
\begin{subequations}\label{met_efet_mag}
    \begin{eqnarray}
  \bar{A}(r) &=& \frac{A(r)}{{\cal L}_{F}},
\\
    \bar{B}(r)&=& \frac{1}{A(r){\cal L}_{F}},
    \\
    \bar{C}(r)&=& \frac{r^2}{{\cal L}_{F}+2F{\cal L}_{FF}}.
    \end{eqnarray}
\end{subequations}

Thus, since we are developing solutions with a matter source coupled to NLED, it is essential to consider the effective metric~\eqref{g_efet}, as particle trajectories are governed by this metric. Furthermore, given that our solutions involve a magnetic charge, we will specifically use the metric described by Eqs.~\eqref{met_efet_mag} to determine the shadow radius.
In the following, we present the equations that enable the computation of the shadow radius, incorporating the effective metric framework and accounting for the influence of the magnetic charge on the geometry of the black hole spacetime.

\subsection{Shadow radius}

The limit of the shadow observed by a distant observer is determined by the critical curve that separates the scattered paths from the captured paths. This critical curve is called the photon sphere ($r_{ps}$), and the solution of the following equation gives us this critical curve
\begin{equation}
\Bar{A}(r_{ps})\,\Bar{C}^\prime(r_{ps})=\Bar{A}^\prime(r_{ps})\,\Bar{C}(r_{ps}).\label{rph}
\end{equation}

This allows us to determine the radius of the shadow of the black hole for a distant observer using the following expression
\begin{equation}
r_{sh}=r_{ps}\sqrt{\frac{\bar{A}(r_{0})}{\bar{A}(r_{p})}}.\label{r_shadow}
\end{equation}

So now we will use the data from Table \ref{tab:my-table} to constrain the parameter $\beta$ of our model based on Eq. \eqref{r_shadow}.
In this way, we will check whether the shadow radius determined in our model matches that observed by the EHT for Sgr A*. 

\begin{table}[ht!]
\centering
\resizebox{\linewidth}{!}{%
\def\arraystretch{1.4}
\begin{tabular}{cccc}
\hline \hline
\multicolumn{4}{c}{Parameter values}                                                                               \\ \hline
\hspace{1 mm}Survey\hspace{1 mm} & \hspace{1 mm}$M (\times 10^6 M_{\odot})$\hspace{1 mm} & \hspace{1 mm}$D$\hspace{1 mm} (kpc) & \hspace{1 mm}Reference\hspace{1 mm} \\ 
Keck                             & $3.951 \pm 0.047$                                     & $7.953 \pm 0.050 \pm 0.032$         & {\cite{Do:2019txf}} \\ 
VLTI                             & $4.297 \pm 0.012 \pm 0.040$                           & $8.277 \pm 0.009 \pm 0.033$         & {\cite{GRAVITY:2020gka}} \\ \hline
\end{tabular}%
}
\caption{Sgr A* Mass and distance.}
\label{tab:my-table}
\end{table}

Geometrically, our spacetime is not asymptotically flat, so the Arnowitt-Deser-Misner (ADM) asymptotic mass \cite{Arnowitt:1962hi} is not well defined for an observer at radial infinity. Therefore, we cannot relate the parameter ``$M$'' that appears in our metric to the mass. For this reason, we specify the shadow radius in units of the local mass ``$M$'', as is the case in \cite{Vagnozzi:2022moj}.
Finally, we calculate the shadow radius of this model so that it is described in terms of $r_O$. We will therefore consider an observer located at $r_O\sim8$ Mpc \cite{Do:2019txf}.

\subsubsection{Constraining the $\beta$ parameter }

\begin{figure}[htbp]
   \includegraphics[scale=0.75]{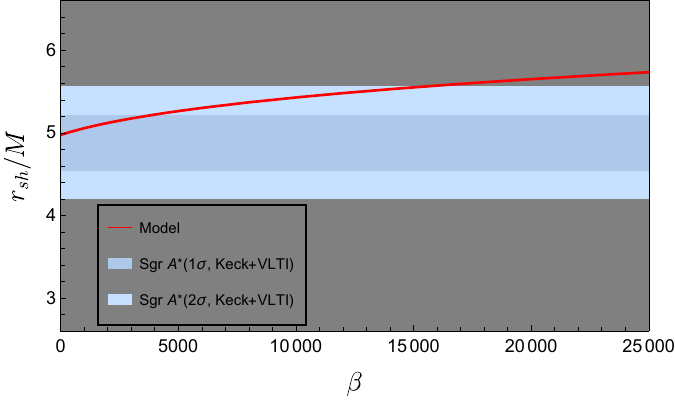}
    \caption{The shadow radius $r_{sh}$ for our black hole model (solid red curve), described by the metric function given in Eq.\eqref{A} with power $p=2$, was calculated according to Eq.~\eqref{r_shadow} for an observer located at $r_O \sim 8000 \ \text{kpc}$. The shaded regions in blue and light blue represent the EHT constraints for the shadow radius of Sgr A* at confidence levels of $1\sigma$ and $2\sigma$, respectively. The gray regions, beyond the $2\sigma$ values, are excluded by the EHT observations. The dark blue and light blue areas represent regions consistent with the EHT observations of Sgr A*, corresponding to confidence levels of $1\sigma$ and $2\sigma$, respectively. }
    \label{figshadowp2}
\end{figure}

In Fig. \ref{figshadowp2} we analyze the behavior of the shadow radius $r_{sh}$ of our model for the metric function \eqref{A} with power $p=2$, i.e. Eq. \eqref{Ap2}. We consider the shadow size of the black hole Sgr A* as a function of the parameter $\beta$ and follow the uncertainties given in Eqs. \eqref{eq:1sigma} and \eqref{eq:2sigma}. The assumed values of the constants are: $M=1$, $q=0.5 M$, $\alpha=0.5$, $f_0=\Lambda/2$, $f_1=10^{-60}$, and $\Lambda=10^{-41}$.
We observe that for values of $\beta$ close to $1.6 \times 10^4$ the shadow radius starts to deviate from the expected value for Sgr A*. For values of $\beta$ below this limit, as shown in Fig. \ref{figshadowp2}, the shadow radius remains within the expected range according to the EHT observations. In Fig. \ref{figshadowp2} we also notice that increasing $\beta$ leads to a larger shadow radius in our model. We also note that the model does not directly approach this limit for values close to $\beta \sim 0$, regardless of whether the power is $p=2$, $p=4$ or $p=6$.

As already mentioned, we have observed in the metric function resulting from the Eq. \eqref{A} the presence of an effective cosmological constant defined in the Eq. \eqref{Lambda} and expressed as $\Lambda_{\rm eff} = 2f_0(1 + 2\beta)$. By analyzing the maximum value of $\beta$ that still keeps the curve of the shadow radius within the confidence level of the observations, i.e. $\sim 1.6 \times 10^4$, we derive a value for the effective cosmological constant of $\Lambda_{\rm eff} \sim 3 \times 10^{-37}$, for the case with $p=2$.

\begin{figure}[t!]
   \includegraphics[width=\linewidth]{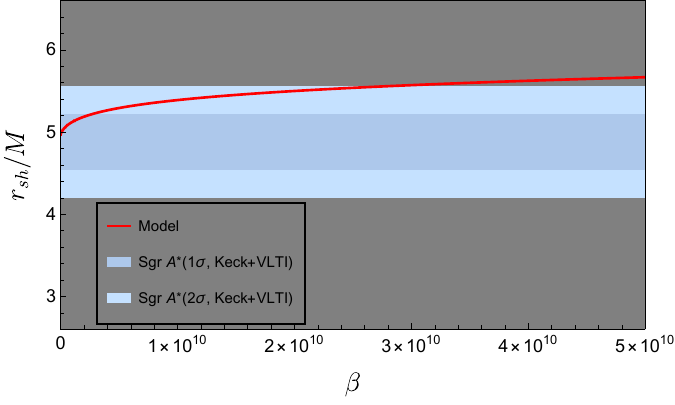}
    \caption{The shadow radius $r_{sh}$ for our black hole model (solid red curve), described by the metric function given in Eq.\eqref{A} with power $p=4$, was calculated according to Eq.\eqref{r_shadow} for an observer located at $r_O \sim 8000 \ \text{kpc}$. The shaded regions in blue and light blue represent the EHT constraints for the shadow radius of Sgr A* at confidence levels of $1\sigma$ and $2\sigma$, respectively. The gray regions, beyond the $2\sigma$ values, are excluded by the EHT observations. The dark blue and light blue areas represent regions consistent with the EHT observations of Sgr A*, corresponding to confidence levels of $1\sigma$ and $2\sigma$, respectively.}
    \label{figshadowp4}
\end{figure}

Let us now analyze the behavior of the shadow radius $r_{sh}$ for the metric function \eqref{Ap4}, compared to the shadow of Sgr A*, as a function of the parameter $\beta$, as shown in Fig. \ref{figshadowp4}. The values of the constants are the same as those used for the case where the power is $p = 2$. We note that for a value of $\beta$ approaching $\sim 3 \cdot 10^{10}$, the shadow radius of this model starts to deviate from the constraints imposed by the shadow radius estimated for Sgr A*. On the other hand, for values of $\beta$ smaller than this, the shadow radius of our model remains within the limits consistent with the EHT observations for Sgr A*, as shown in Fig. \ref{figshadowp4}. Since the model covers the range of values $0 \leq \beta \lesssim 3 \times 10^{10}$, we use the maximum value $\beta_{\text{max}} \sim 3 \times 10^{10}$ to calculate the effective cosmological constant. In this case, we obtain $\Lambda_{\rm eff} \sim 3 \times 10^{-31}$.

\begin{figure}[t!]
   \includegraphics[width=\linewidth]{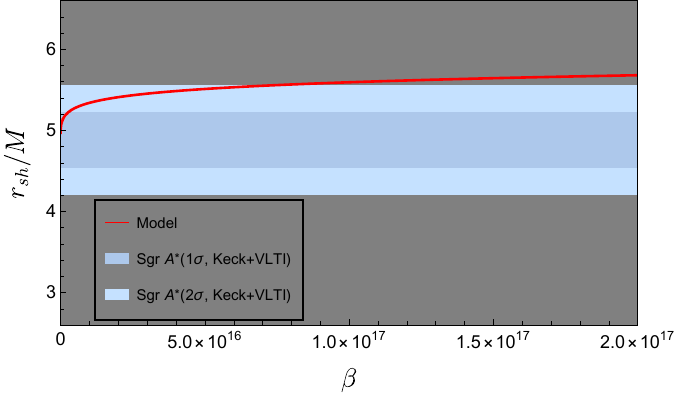}
    \caption{The shadow radius $r_{sh}$ for our black hole model (solid red curve), described by the metric function given in Eq.\eqref{A} with power $p=6$, was calculated according to Eq.\eqref{r_shadow} for an observer located at $r_O \sim 8000 \ \text{kpc}$. The shaded regions in blue and light blue represent the EHT constraints for the shadow radius of Sgr A* at confidence levels of $1\sigma$ and $2\sigma$, respectively. The gray regions, beyond the $2\sigma$ values, are excluded by the EHT observations. The dark blue and light blue areas represent regions consistent with the EHT observations of Sgr A*, corresponding to confidence levels of $1\sigma$ and $2\sigma$, respectively.}
    \label{figshadowp6}
\end{figure}

As in the previous cases, we have also analyzed the behavior of the shadow radius $r_{sh}$ for the power value $p = 6$, as illustrated in Fig. \ref{figshadowp6}. The values of the constants used are the same as in the previous cases. We note that when the value of the parameter $\beta$ approaches $\sim 7.8 \times 10^{16}$, the shadow radius of the developed model starts to deviate from the constraints imposed by the shadow radius limit estimated for Sgr A*. On the other hand, for smaller values of $\beta$, the shadow radius of our model remains consistent with the observations of Sgr A*. Finally, for the maximum value of $\beta$, $\beta_{\text{max}}$, we calculate the effective cosmological constant, which becomes $\Lambda_{\rm eff} \sim 10^{-24}$. If we compare this value with the values of the effective cosmological constant for models with lower powers obtained by assuming the maximum values of $\beta_{\text{max}}$, we come to the conclusion that the value of the effective cosmological constant increases with increasing $\beta$.

We also check the radius of the shadow of the black hole for the cases $f_0 = 0$ and $f_0 < 0$. In the case of $f_0 < 0$, the effective cosmological constant becomes negative. However, we did not find any significant differences to the results shown in Figs. \ref{figshadowp2}, \ref{figshadowp4} and \ref{figshadowp6}, which is why we decided not to include these approaches in the present work.

In addition, we have adopted a more general approach to estimate the value of the parameter $\beta$. We numerically analyze the values described by Eq. \eqref{function}, which depend directly on the scalar $R$ and the coupling of $\beta$ with the trace $T$ of the energy-momentum tensor. From this, we determine the order of magnitude associated with the parameter $\beta$ with respect to $R$ and $T$ and verify that it corresponds to the value resulting from the shadow radius we analyzed. For the values of the constants we used, we found that the term involving $\beta T$ has an order of magnitude of $10^{-41}$, which appears to agree with the behavior we observed for the shadow radius of our model, indicating high values for the parameter $\beta$. On the other hand, when we analyze only the numerical value for the scalar $R$, we find an order of magnitude of $10^{-41}$.

\subsubsection{Constraining the $\alpha$ parameter }

\begin{figure}[t!]
   \includegraphics[width=\linewidth]{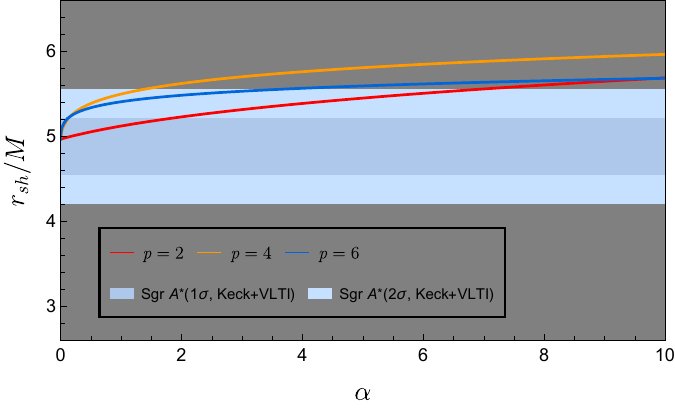}
    \caption{The shadow radius $r_{sh}$ for our black hole model (solid red, yellow and blue curve), described by the metric function given in Eq.\eqref{A} with power $p=2,4,6$, was calculated according to Eq.\eqref{r_shadow} for an observer located at $r_O \sim 8000 \ \text{kpc}$. The shaded regions in blue and light blue represent the EHT constraints for the shadow radius of Sgr A* at confidence levels of $1\sigma$ and $2\sigma$, respectively. The gray regions, beyond the $2\sigma$ values, are excluded by the EHT observations. The dark blue and light blue areas represent regions consistent with the EHT observations of Sgr A*, corresponding to confidence levels of $1\sigma$ and $2\sigma$, respectively.}
    \label{figshadowalpha}
\end{figure}

Similar to the procedure developed in the previous section, we now constrain the parameter $\alpha$ based on the data obtained by the Event Horizon Telescope (EHT) for Sgr A*, as illustrated in Fig. \ref{figshadowalpha}. We analyze the behavior of the shadow radius $r_{sh}$ for the metric function of our model described by Eq. \eqref{A}, considering the same values for the constants $M$, $q$, $f_0$, $f_1$, and $\Lambda$ that were used in the constraint of the parameter $\beta$. We investigate three different scenarios corresponding to the power values $p = \{2, 4, 6\}$.  

Regarding the parameter $\beta$, we use the values obtained from the analysis in the previous section. More specifically, for each value of $p = \{2, 4, 6\}$, we assume the corresponding values of $\beta$: $\beta = 10^3$, $\beta = 10^{10}$, and $\beta = 10^{16}$, respectively. Based on this, we compute the shadow radius using Eq. \eqref{r_shadow} from the models given by Eqs. \eqref{Ap2}, \eqref{Ap4}, and \eqref{Ap6}. The size of the shadow of the black hole Sgr A* is then represented as a function of the parameter $\alpha$ in Fig. \ref{figshadowalpha}, where three different curves are shown: red, yellow, and blue, corresponding to $p=2$, $p=4$, and $p=6$, respectively. 

For the case where $p=2$ and $\beta= 10^3$, represented by the red curve, the shadow radius begins to exceed the limits set by EHT observations when the parameter $\alpha$ approaches the value 7. For smaller values of $\alpha$, the shadow radius remains within the allowed range, as illustrated in Fig. \ref{figshadowalpha}.

For $p=4$ and $\beta= 10^{10}$, the shadow radius starts to deviate from the constraints imposed by the estimated value for Sgr A* when $\alpha \sim 1.4$. However, for values of $\alpha$ smaller than this, the shadow radius of our model remains consistent with the EHT observational limits for Sgr A*, as indicated by the yellow curve in Fig. \ref{figshadowalpha}.

Finally, for $p=6$ and $\beta= 10^{16}$, the shadow radius begins to deviate from the expected value for Sgr A* at approximately $\alpha \sim 3.8$. For values of $\alpha$ below this limit, as shown in Fig. \ref{figshadowalpha}, the shadow radius remains in agreement with the EHT observations.

Furthermore, Fig. \ref{figshadowalpha} shows that increasing the parameter $\alpha$ leads to an increase in the shadow radius, a trend similar to the behavior observed when constraining the parameter $\beta$. This pattern is maintained across different values of $p$.

\subsubsection{Constraining the  $p$ parameter}

\begin{figure}[t!]
   \includegraphics[width=\linewidth]{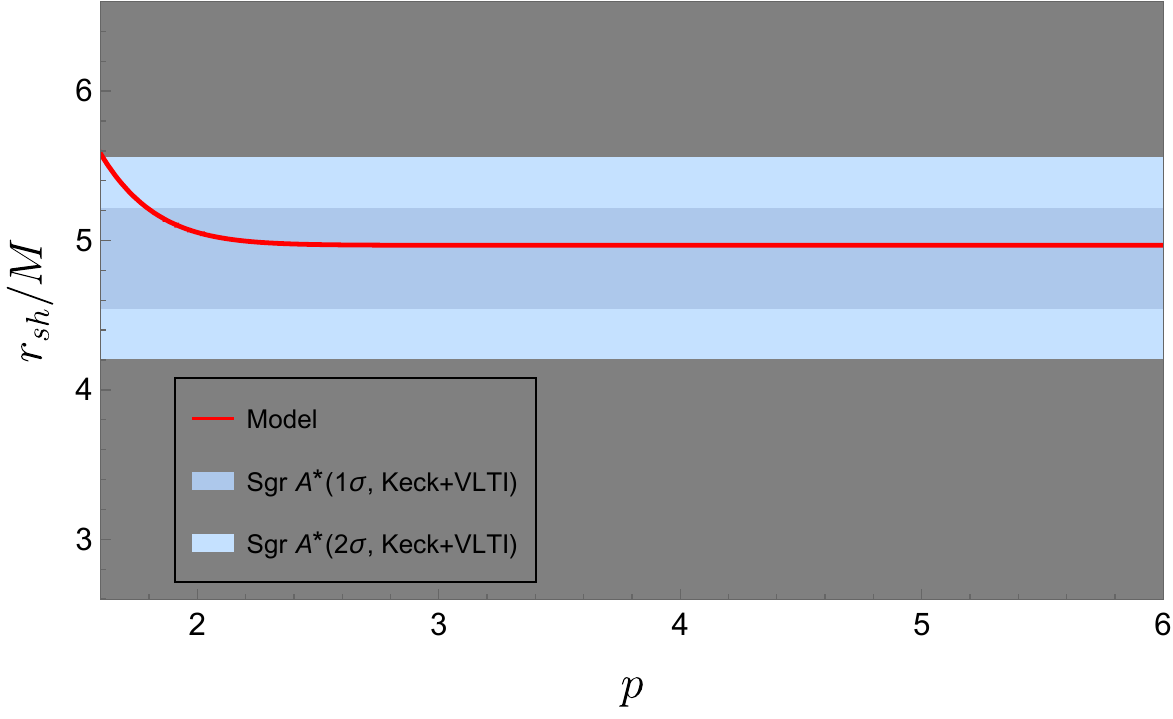}
    \caption{The shadow radius $r_{sh}$ for our black hole model (solid red, yellow and blue curve), described by the metric function given in Eq.\eqref{A}, was calculated according to Eq.\eqref{r_shadow} for an observer located at $r_O \sim 8000 \ \text{kpc}$. The shaded regions in blue and light blue represent the EHT constraints for the shadow radius of Sgr A* at confidence levels of $1\sigma$ and $2\sigma$, respectively. The gray regions, beyond the $2\sigma$ values, are excluded by the EHT observations. The dark blue and light blue areas represent regions consistent with the EHT observations of Sgr A*, corresponding to confidence levels of $1\sigma$ and $2\sigma$, respectively.}
    \label{figshadowp}
\end{figure}

Finally, as in the two previous cases, we constrain the parameter $p$. To achieve this, we use the same values for the constants and set $\alpha=0.5$ and $\beta=10^3$. The obtained results provide strong justification for the values of $p$ that were assumed in the development of our solutions. When analyzing the behavior of the shadow radius of our model within the range $p \in [2,6]$, we clearly observe that its values remain entirely within the range compatible with the EHT observations for the black hole Sgr A*. This result validates our choices in constructing the solutions presented in Section \ref{sec3}. Fig. \ref{figshadowp} illustrates this analysis.

\section{Summary and Future Perspectives}\label{sec:concl}

\subsection{Outline of the main results }

In this paper, we investigated spherically symmetric black hole solutions within the framework of $f(R,T)$ gravity, where $R$ represents the Ricci scalar and $T$ is the trace of the energy-momentum tensor. We explored these solutions by coupling the $f(R,T)$ field equations, as given by Eq.~\eqref{EqM}, with the matter content described by nonlinear electrodynamics (NLED). To proceed with the analysis, we assumed that the electromagnetic scalar component is solely determined by the magnetic charge.
To develop our solutions, we first considered a linear $f(R,T)$ function, as described by Eq.~\eqref{function}, and assumed a Lagrangian of the form with an arbitrary power to facilitate our calculations. From the components of the field equations, specifically Eqs.~\eqref{EqF00}--\eqref{EqF22}, we derived the metric function through integration, as detailed in Eq.~\eqref{A}. Subsequently, we specialized this metric function to examine specific solutions corresponding to defined powers, such as $p = 2$, $p = 4$, and $p = 6$.

After determining the metric functions corresponding to these powers, we proceeded to investigate the existence of event horizons by obtaining numerical solutions. To develop these solutions, we fixed specific values for certain constants, allowing us to separately analyze the critical charge and the critical mass of the system.
We observeded that the metric functions derived for the powers $p=2$, $p=4$, and $p=6$ exhibited similar qualitative behavior in the numerical solutions for the horizons. Consequently, the solutions discussed in this study are based on the metric function \eqref{Ap6}. In the first model, we solved the equations \eqref{rH} and \eqref{der_a} simultaneously to explore the horizon structure.
For the critical mass $M_c$, we observed the following behavior: in one of the solutions, when the mass $M > M_c$, four horizons are present. At the critical mass $M = M_c$, three horizons are observed, while for $M < M_c$, only two horizons appear.
Similarly, for the critical charge $q_c$, the solution revealed that for $q > q_c$, there is only one horizon. At $q = q_c$, two horizons exist, one of which is degenerate. For $q < q_c$, we present three curves that illustrate a scenario in which up to three horizons can emerge, highlighting a particularly interesting case with a maximum of three horizons.

We also calculated and analyzed the Kretschmann scalar for the solutions corresponding to the powers $p=2$, $p=4$, and $p=6$. Our investigation revealed that, for all these powers, the Kretschmann scalar exhibits a divergence for very small values of the radial coordinate $r$, indicating a potential singularity at the origin. However, as the radial coordinate $r$ increases, all the Kretschmann scalars display regular behavior, and specifically, they tend to a constant value given by $8 \Lambda_{\text{eff}}^2 / 3$ in the large-$r$ limit. This regularity suggests that the solutions asymptotically approach a de Sitter-like spacetime at large distances.
Furthermore, in Section \ref{top2}, we presented two different approaches used to derive the analytical form of the Lagrangian in terms of the electromagnetic field strength $F$. After calculating the components of the Lagrangian, namely ${\cal L}_F(r)$ and ${\cal L}_{FF}(r)$, from the equations of motion (Eqs. \eqref{EqF00} and \eqref{EqF22}), we observed that regardless of the specific consistency relation (either \eqref{RC} or \eqref{RC2}) initially chosen for the calculation, the resulting analytical form of the Lagrangian ${\cal L}_{\text{NLED}}(F)$ remains invariant. 
To verify this result, we employed the metric function described by \eqref{A}. Notably, this outcome holds true irrespective of the metric function selected, further confirming the robustness and consistency of our derived Lagrangian.

Furthermore, we observed that the Lagrangian given by Eq. \eqref{Lag_F}, which was derived using the last formalism, retains its non-linear character even when the constant $\alpha$ is set to zero. It is important to recall that this constant $\alpha$ was initially included in the Lagrangian \eqref{L}. In the absence of $\alpha$, the non-linearity in the Lagrangian persists, but this non-linearity is now driven by the constant $\beta$, which is directly coupled to the trace of the energy-momentum tensor within the function \eqref{function} that we proposed. 
Even in the case where the term involving $\beta$, which appears as an exponent of the field strength $F$ in the Lagrangian, is removed, the Lagrangian continues to exhibit non-linear behavior. This is due to the remaining term in the Lagrangian that is proportional to $F$ and governed by the constant $\alpha \neq 0$. Specifically, the non-linearity stems from the term with an arbitrary power $p$ of the electromagnetic field strength $F$, which introduces the desired non-linear effects.
Notably, while the presence of these non-linearities prevents the Lagrangian from being of the simple linear form typical of Maxwell electromagnetism, it is still possible to recover the linear Maxwell case under specific conditions. This can be achieved by imposing the appropriate constraints on the constants $\alpha$ and $\beta$, which would yield the standard linear electromagnetic theory. However, even in this limit, the system would not revert to standard GR. Instead, we would continue working within the framework of $f(R,T)$ theory, which encapsulates the effects of an effective cosmological constant, reflecting the inherent modifications to the gravitational sector due to the $f(R,T)$ formulation. This underscores the key distinction between the linear Maxwell theory and the more general $f(R,T)$ theory, where even the linear limit retains the modifications introduced by the presence of $\beta$ and the trace of the energy-momentum tensor.

Finally, we investigated the behavior of the black hole shadow predicted by our model and compared it with the shadow size of the supermassive black hole at the center of our galaxy, Sagittarius A* (Sgr A*), as observed by the Event Horizon Telescope (EHT). The goal of this analysis was to constrain the parameter $\beta$, $\alpha$ and $p$ in our model, which plays a crucial role in shaping the shadow's characteristics. Specifically, we examined the influence of $\beta$ on the shadow radius for three different values of the power $p$: $p=2$, $p=4$, and $p=6$.
In the scenarios we analyzed, we observed that the parameter $\beta$ takes large values, and we determined the ranges for the shadow radius of our black hole, which were constrained by the EHT observations of Sgr A*. The estimated intervals for $\beta$ corresponding to each value of $p$ are as follows:
$0 \leq \beta_{p=2} \lesssim 1.6 \times 10^{4}$, $0 \leq \beta_{p=4} \lesssim 3 \times 10^{10}$, and $0 \leq \beta_{p=6} \lesssim 7.8 \times 10^{16}$.
These intervals progressively increase as the power $p$ increases, highlighting a direct relationship between the value of $p$ and the allowable range for $\beta$. Moreover, we observed that as $\beta$ increases, the radius of the black hole’s shadow also increases for all values of $p$. This indicates that the shadow's size is highly sensitive to the value of $\beta$, which is consistent across all considered powers.

Similar to the procedure that was developed to constrain the $\beta$ parameter, we also constrain and analyze the impact of the $\alpha$ parameter on the behavior of the black hole's shadow radius for three different values of the power $p = (2,4,6)$. For the $\beta$ parameter, we use the values that are within the margin of constraint we obtained. More specifically, we use: 
$\beta_{p=2} = 10^{3}$, $\beta_{p=4} = 10^{10}$, $\beta_{p=6} = 10^{16}$.
In this context, we note that the intervals for the radius of the shadow of our black hole, limited by the EHT observations of Sgr A*, correspond to the following values of $\alpha$ for each value of $p$:
$0 \leq \alpha_{p=2} \lesssim 7$, $0 \leq \alpha_{p=4} \lesssim 1.4$, $0 \leq \alpha_{p=6} \lesssim 3.8$.
These intervals show that, as $p$ increases, the value of $\alpha$ decreases. More specifically, the $\alpha$ parameter reaches its lowest value, within observational limits, when $p = 4$, and then increases again as $p$ approaches 6. 
In addition, we observe that as the $\alpha$ parameter increases, the radius of the black hole's shadow also increases, for all the values of $p$ analyzed.

Finally, to justify the choice of values for $p = (2, 4, 6)$ used in this manuscript, we also analyzed the influence of the parameter $p$ on the shadow radius of our model. To do this, we used the values $\beta = 10^3$ and $\alpha = 0.5$. As a result, we found that by restricting the parameter $p$, the shadow radius remains completely within the values corresponding to the EHT observations for Sgr A*.

Additionally, we calculated the effective cosmological constant associated with the model for the maximum values of $\beta$ within the specified intervals ($\beta_{\text{max}}$ for each power $p = 2$, $4$, and $6$), ensuring that the shadow radius does not exceed the confidence limits set by the Sgr A* observations. We found that as $\beta$ increases, the effective cosmological constant also increases, following the same trend as $\beta$ for the different values of $p$. This suggests a close connection between the parameter $\beta$ and the effective cosmological constant, which further emphasizes the significance of $\beta$ in modifying the structure of the black hole's shadow, as well as its relation to the cosmological dynamics within the context of our model.

\subsection{Future Perspectives}

In this study, we have explored the interplay between nonlinear electrodynamics and $f(R,T)$ gravity, providing new insights into how modifications to the matter source influence the geometry and physical properties of black hole spacetimes. The solutions derived in this framework offer a broader perspective on black hole physics, particularly in the context of modified theories of gravity, highlighting the complex interplay between gravitational and electromagnetic fields. These findings pave the way for a deeper understanding of black hole solutions within more general and extended gravitational theories.
Despite the wealth of existing literature on this formulation, our work opens several avenues for future research. One promising direction is the application of $f(R,T)$ gravity to investigate black hole thermodynamics, including the study of thermodynamic stability and the possible implications for the laws of black hole mechanics. Furthermore, the analysis of quasi-normal modes within this framework could reveal important features of black hole oscillations, providing additional clues about the underlying structure of these exotic objects.

Additionally, we plan to further examine the shadow of black holes in $f(R,T)$ gravity, an essential observational signature that can constrain various model parameters. Our results have already provided an initial comparison with the shadow of Sgr A*, and future work may refine these results with higher precision observations, offering further insight into the role of modified gravity in shaping the black hole's observable properties.
The study of gravitational waves and quasi-periodic oscillations in the context of $f(R,T)$ gravity is another exciting avenue. As gravitational wave detectors become more sensitive, the ability to probe black holes in modified gravity theories could open new windows into both fundamental physics and astrophysical observations. Moreover, extending our analysis to rotating black holes within the $f(R,T)$ framework, such as those described by Kerr metrics, would provide a more complete understanding of the role of nonlinear electrodynamics in curved spacetimes.

Finally, we are also interested in the application of $f(R,T)$ gravity to more exotic scenarios, such as black-bounce solutions, which could offer novel alternatives to the traditional black hole model, particularly in the context of regularizing singularities. 
In summary, the $f(R,T)$ theory provides a rich and versatile framework for exploring black hole solutions, and its potential applications extend across a wide range of topics in gravitational physics. Our future work will continue to explore these areas, furthering our understanding of the dynamics of black holes in modified gravitational theories and their observational signatures.


\acknowledgments{

We would like to thank our collaborators Henrique A. Vieira and Luís F. Dias da Silva for the subtle modifications in the Mathematica code that allowed us to develop the results on the shadow radius.
MER thanks Conselho Nacional de Desenvolvimento Cient\'ifico e Tecnol\'ogico - CNPq, Brazil, for partial financial support. This study was financed in part by the Coordena\c{c}\~{a}o de Aperfei\c{c}oamento de Pessoal de N\'{i}vel Superior - Brasil (CAPES) - Finance Code 001.
FSNL acknowledges support from the Funda\c{c}\~{a}o para a Ci\^{e}ncia e a Tecnologia (FCT) Scientific Employment Stimulus contract with reference CEECINST/00032/2018, and funding through the research grants UIDB/04434/2020, UIDP/04434/2020 and PTDC/FIS-AST/0054/2021.}



\end{document}